\documentclass{an}
\usepackage{graphicx}
\usepackage{subfigure}
\usepackage{times}
\usepackage{fancyhdr}
\begin{document}

\title{10 New Very Low Mass Close Binaries Resolved in the Visible}

\author{Nicholas M. Law \inst{1}, Simon T. Hodgkin \inst{1}, Craig D. Mackay \inst{1}, John E. Baldwin \inst{2}}
\institute{Institute of Astronomy, University of Cambridge, Madingley Road, Cambridge CB3 0HA \and Cavendish Astrophysics Group, Cavendish Laboratory, Madingley Road, Cambridge}

\date{Received 16 October 2005; accepted 1 November 2005; published online 15 December 2005}

\abstract{We present preliminary results from the first part of the LuckyCam late M-dwarf binarity survey. We survey a sample of 48 nearby ($<$40 pc) and red (M5-M9) stars with the novel high angular resolution visible light imaging technique Lucky Imaging, in only 8 hours of 2.5m telescope time. We discover 10 new binaries; although the survey is sensitive to brown dwarf companions none are detected. The orbital radius distribution of the newly discovered binaries broadly matches that of previous detections by other groups, although we do discover one wide binary at $\sim$40AU.
}

\correspondence{nlaw@ast.cam.ac.uk}

\maketitle

\section{Introduction}
This short paper presents preliminary results from our \mbox{LuckyCam} high angular resolution binarity survey of nearby ($<$40 pc) and red (M5-M9) stars. LuckyCam is a novel imaging system designed and built by our group that gives near diffraction limited images in the visible with 2.5m class telescopes.

\section{The Sample}
\begin{figure}
  \centering
  \resizebox{\columnwidth}{!}
   {
     \includegraphics{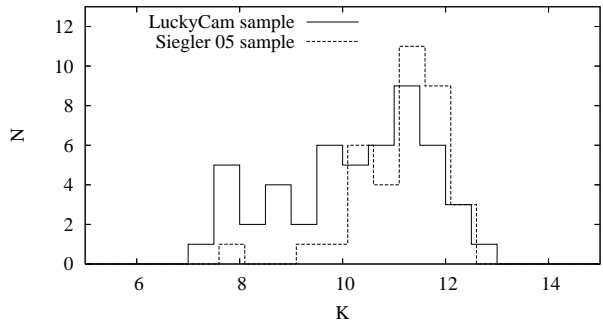}
   }
   \caption{\footnotesize The V-K colour distribution and approximate spectral class range of our sample. Spectral classes are estimated from the V-K relations given in Leggett 1992. For comparison we show the 36 star sample of Siegler et al. 2005, observed with Gemini North, VLT, Keck II and Subaru. }
   \label{FIG:k_hist}
\end{figure}

The V-K colour distribution and approximate spectral class range of our sample is shown in figure \ref{FIG:k_hist}. Targets were selected with the following criteria:

\begin{enumerate}
\item{$\rm{PM < 0.15''/year}$; this selects only nearby targets and thus eliminates giants which may otherwise contaminate our colour selected sample.}
\item{$\rm{V-K > 6}$; thus selecting M5 and later stars (Leggett 1992).}
\item{$\rm{D < 40pc}$; based on absolute magnitudes estimated from V-K colours and the photometric parallax calibrations described in Leggett 1992.}
\item{$\rm{m_{SDSS i'} < +15.5}$; Lucky Imaging requires at least an $\rm{m_{SDSS i'} = 15.5}$ guide star for full performance. All targets serve as their own guide star.}
\item{We removed all stars that to our knowledge have been previously observed at high angular resolution.}

\end{enumerate}

The resulting list consisted of over 200 targets, 47 of which were observed in June 2005. We chose the observed targets on the basis of magnitude and, in 16 cases, association with X-ray sources. An analysis of the sample bias and completeness will be required to obtain good constraints on the binary fractions of these targets; we here present these preliminary results as an initial validation of our detection techniques.

\section{Observations and the Lucky Imaging Technique}
Our Lucky Imaging system (LuckyCam) takes a sequence of images at $>$10 frames per second using a very low noise L3CCD based conventional camera. Each frame experiences different atmospheric effects. To construct a high resolution long exposure image we select, align and co-add only those frames which meet a quality criterion. Because Lucky Imaging is a passive technique data can be taken as soon as telescope pointing is completed allowing extremely efficient observing. The technique is described in more detail in Law et al. 2005. 

We have found that the system can correct images in the visible to close to the diffraction limit of the Nordic Optical Telescope (NOT) in good seeing using reference stars as faint as i=15.5. Even in poor seeing Lucky image selection strategies deliver an improvement in image resolution of as much as a factor of four.

We imaged all 47 targets in only 8 hours of on-sky time during the nights of 3-6 June 2005, on the 2.56m NOT. Each target was imaged for 100 seconds in both the SDSS i' and z' filters. Seeing values ranged from 0.5 arcseconds to 1.2 arcseconds with a median of 0.8''; the final FWHM resolutions were in all cases better than 0.15''.

\section{Discovered binaries}
\begin{figure}
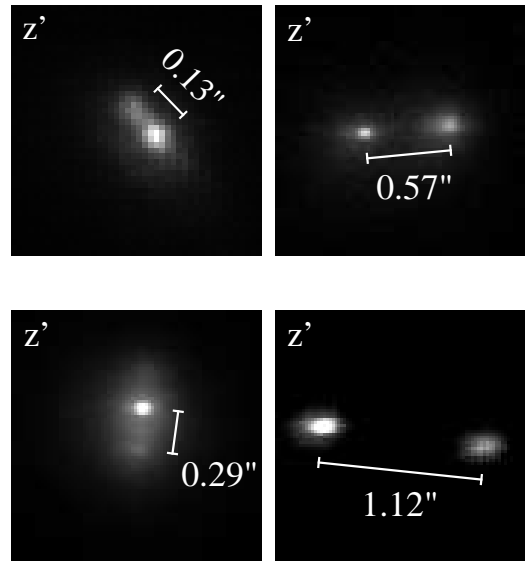

  \centering
	\subfigure{\includegraphics[width=0.40\columnwidth]{1_z.eps}}
	\subfigure{\includegraphics[width=0.40\columnwidth]{2_z.eps}}
	\subfigure{\includegraphics[width=0.40\columnwidth]{3_z.eps}}
	\subfigure{\includegraphics[width=0.40\columnwidth]{4_z.eps}}
   \caption{\footnotesize SDSS z' images of some of the discovered binaries; 10\% selection (1\% for top left) from 3000 frames.}
   \label{FIG:binaries}
\end{figure}

We detected 10 new very close binaries with separations between 0.13'' and 1.5''. The distribution of orbital radii is shown in figure \ref{FIG:sep_hist}; example images are shown in figure \ref{FIG:binaries}. In all 48 20x20" full fields only one red background object was detected. Limiting our binary search to 2 arcseconds radius from each target star, we would thus expect only 0.03 spurious associations in our dataset. We therefore (on a preliminary basis) conclude that all the detected close candidates are physically associated.

\section{Discussion and conclusions}
Because the sample presented in this paper is inhomogeneous we feel it would be premature to calculate a binary fraction on the basis of our detections; this will be addressed in further work.

Although we are sensitive to very faint companions (i'=20 magn. at 1 arcsec, $\Delta$m=5, in the brown dwarf regime), none of the detected companions are more than 2 magnitudes fainter than the primary. This suggests that, as has been found by previous authors, brown-dwarf companions to M-dwarfs are very rare at separations wider than a few AU.

\begin{figure}
  \centering
  \resizebox{\columnwidth}{!}
   {
	\includegraphics{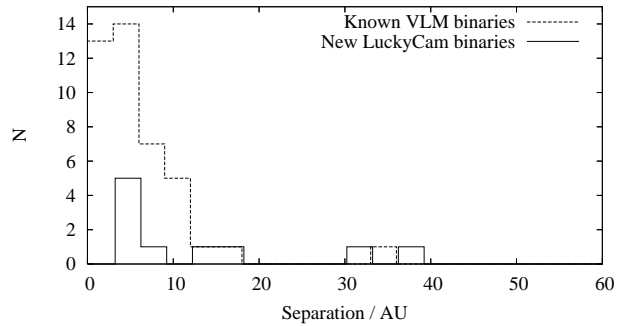}
   }
   \caption{\footnotesize The distribution of orbital radii of the new binaries compared to that of previously known VLM binary systems collated in Siegler et al. 2005.}
   \label{FIG:sep_hist}
\end{figure}

The distribution of orbital radii (figure \ref{FIG:sep_hist}) is consistent with that found in previous M-dwarf surveys, with a peak at around 4 AU and no companions found at larger radii than 40AU. However, we note that two of our new binaries fall near a previous single detection at about 35 AU. It is not yet clear  whether these wider systems are in the high separation tail of the  orbital radius distribution, or part of a distinct population.

\acknowledgements
NML acknowledges support from the UK Particle Physics and Astronomy Research Council (PPARC). Based on observations made with the Nordic Optical Telescope, operated on the island of La Palma jointly by Denmark, Finland, Iceland, Norway, and Sweden, in the Spanish Observatorio del Roque de los Muchachos of the Instituto de Astrofisica de Canarias.

\end{document}